# Novel insulator to metal transition and superconductivity in Sr-doped $La_2CuO_4$

**Pei Herng Hor**[1] **and Young Hoon Kim**[2]

[1] Department of Physics and Texas Center for Superconductivity, University of Houston, Houston, TX 77204-5002, USA
[2] Department of Physics, University of Cincinnati, Cincinnati, OH 45221-0011, USA

**Abstract**
We have studied the charge dynamics of Sr-doped $La_2CuO_4$ at hole concentration $p = 0.07$. We observed a clear indication of the competition between two new characteristic collective modes of the Wigner lattice orders suggested earlier: one corresponds to a $T_c = 15$ K superconductivity and the other to $T_c = 30$ K. Our results naturally explain the recent observation of metallic transport in extremely low (1%) Sr-doped $La_2CuO_4$ and establish a novel insulator to metal transition mechanism and superconductivity in the two-dimensional Wigner lattice ground state. We suggest that the fluctuations of the relative phase between the superconducting condensate and the Wigner lattice order can provide an answer to the unidentified mode in the $c$-axis conductivity that appears only below $T_c$.

## 1. Introduction

The superconducting (SC) order in cuprates originates from the two-dimensional charge dynamics of the doped holes (or electrons) in the $CuO_2$ planes. The undoped cuprates at half-filling exhibit two-dimensional antiferromagnetic order. Away from half-filling, the SC order sets in as the doping concentration ($p$) increases beyond the critical hole concentration $p_c = 1/16$ and a generic phase diagram as a function of $p$ and temperature ($T$) has been constructed using a variety of experimental probes [1]. Therefore, understanding the charge dynamics in the $CuO_2$ planes away from half-filling is particularly relevant. In our recent far-infrared (far-IR) study of the Sr/O co-doped $La_{1.985}Sr_{0.015}CuO_{4+\delta}$ system, we have observed two new collective modes [2]. As a plausible interpretation, we have attributed these modes to the two-dimensional Wigner lattice orders and subsequently showed that high-temperature superconductivity can be naturally realized through the phase coherence transition based on these two-dimensional Wigner lattice order ground states away from half-filling. Since the new collective modes have never been observed in spite of extensive previous far-IR studies and due to the fact that we have excess mobile oxygen atoms in the Sr/O co-doped system, it is highly desirable to study the more conventional pure Sr-doped system to further pin down the nature of these modes and elucidate the role of our proposed Wigner lattice order in the normal state properties.

We carried out detailed transport and far-IR reflectivity measurements of a single-phase polycrystalline $La_{1.93}Sr_{0.07}CuO_4$ sample for the frequency range from $\sim$7 to 5000 cm$^{-1}$. We chose a $p = 0.07$ system, $La_{1.93}Sr_{0.07}CuO_4$, which shows an SC transition at $T_c = 16$ K, in order to directly compare with the previously obtained data of Sr/O co-doped $La_2CuO_4$ near $p_c = 1/16$. We studied the charge dynamics in the $CuO_2$ planes of $La_{1.93}Sr_{0.07}CuO_4$ using a polycrystalline sample for the following two reasons:

(1) there exists a serious problem with uniform distribution of the Sr atoms in preparing single-crystalline samples at low doping concentrations and
(2) far-IR studies of single-crystalline samples are no longer essential because details of the $c$-axis charge dynamics in cuprates are now understood and the $c$-axis far-IR spectral information is well documented.

In this paper, we report the results of our recent charge dynamics studies of 7% Sr-doped $La_2CuO_4$ superconductor ($p = 0.07$ and $T_c = 16$ K). We have interpreted the results based on our proposed two-dimensional Wigner lattice model [2]. We have observed a clear indication of the competition between the p(4 × 4) Wigner lattice (or two-dimensional charge density wave) and the c(2 × 2) lattice order (see figure 3). We found that, as the c(2 × 2) lattice order loses the competition, the $T_c = 16$ K superconductivity results, unlike the case of the $p = 0.07$ Sr/O co-doped counterpart which exhibited $T_c = 26$ K [2]. Our results resolve the problem of the insulator to metal (I–M) transition in cuprates [3] and establish a novel I–M transition mechanism and superconductivity in the two-dimensional Wigner lattice ground state. The issue of the metallic transport in 1% Sr-doped $La_2CuO_4$ for $T > 100$ K [4] can be understood based on our findings. The observed phase collective mode in the SC state provides a channel for the coupling of the $c$-axis-polarized far-IR photon field to the fluctuations of the relative phase between the SC condensate and the Wigner lattice order. We suggest that the unidentified mode at around $\sim$50 cm$^{-1}$ that appears in the $c$-axis conductivity only below $T_c$ [5] arises through this relative phase fluctuation mechanism.

## 2. Experimental details

Far-IR reflectivity was measured at near normal angle of incidence ($\sim$8°) on the sample. A Bruker 113v infrared spectrometer was used. In order to cover the frequency ($\omega$) below 20 cm$^{-1}$, a 75 $\mu$m Mylar beamsplitter and a doped Si-composite bolometer with 1 cm$^2$ active area operating at 2 K was used in conjunction with a parabolic light cone with a 7 mm diameter exit aperture.

The sample temperature was controlled by direct measurement of the temperature from the backside of the sample using a Neocera LTC-11 cryogenic temperature controller. The temperature resolution was $\pm$0.1 K for $T \leqslant 40$ K. Far-IR properties, represented by the real part of the conductivity $\sigma_1(\omega)$ and the real part of the dielectric function $\varepsilon_1(\omega)$, were calculated from a Kramers–Kronig analysis. Spectral resolution was set to 4 cm$^{-1}$ for $\omega$ above 600 cm$^{-1}$, 2 cm$^{-1}$ for 30 cm$^{-1} < \omega < 600$ cm$^{-1}$ and 1 cm$^{-1}$ for $\omega < 30$ cm$^{-1}$.

## 3. Results and discussions

Far-IR and infrared reflectivities of the $La_{1.93}Sr_{0.07}CuO_4$ sample at various $T$ are displayed in figure 1. There appears a sharp minimum in the reflectivity at $\omega \sim 16$ cm$^{-1}$ below which the reflectivity curve resembles that of a free electron gas. Since this sharp minimum occurs at all temperatures, we rule out the possibility that the minimum is the $c$-axis Josephson plasmon edge. Above this $\omega$ edge, three modes are present: one at $\sim 24$ cm$^{-1}$, a weak mode at $\sim 46$ cm$^{-1}$ and the third at $\sim 100$ cm$^{-1}$ in addition to the $c$-axis phonon modes at $\sim 230$ cm$^{-1}$ and $\sim 500$ cm$^{-1}$. Because of the highly featureless, insulating nature of the $c$-axis spectrum even for the optimally doped single crystals, one may safely conclude that the observed spectral features other than the $c$-axis phonons are from the $CuO_2$ planes.

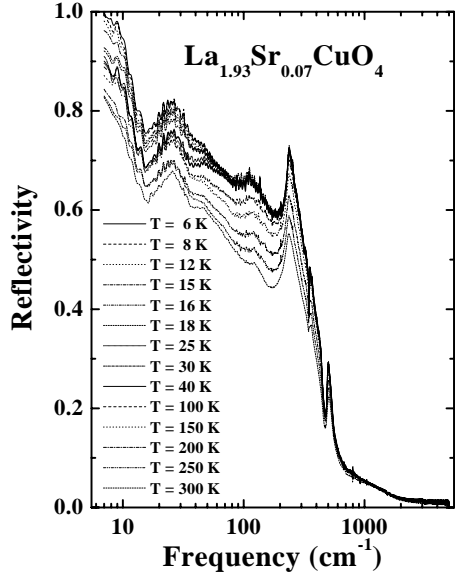

**Figure 1.** Far-IR and infrared reflectivity spectra of polycrystalline $La_{1.93}Sr_{0.07}CuO_4$ at various temperatures.

The Kramers–Kronig-transformation-derived $\sigma_1(\omega)$ and $\varepsilon_1(\omega)$ are shown in figure 2. A number of peaks appear in $\sigma_1(\omega)$. Besides the well known $c$-axis and in-plane phonon modes, a sharp peak at $\sim 24$ cm$^{-1}$ and a broad peak at $\sim 100$ cm$^{-1}$ and weak modes at $\sim 46$ cm$^{-1}$ and at $\sim 70$ cm$^{-1}$ are readily seen and the oscillator strength of these modes show $T$-dependence. A relatively $T$-independent broad excitation peaked at $\sim 1200$ cm$^{-1}$ with an onset at $\sim 400$ cm$^{-1}$ is clearly seen. In addition, there exists a sharp upturn in $\sigma_1(\omega)$ for $\omega < 10$ cm$^{-1}$.

In our recent far-IR studies of $La_{1.985}Sr_{0.015}CuO_{4+\delta}$ samples at the hole concentration $p \sim 0.063$ ($\delta = 0.024$) and $p \sim 0.07$ ($\delta = 0.032$) [2], two two-dimensional Wigner lattice orders were found as the ground state of the $CuO_2$ plane away from half-filling. We observed that, at $p = 0.063$, almost all the holes are used to form a p(4 × 4) square lattice ($p_c = 1/16$) of side $L = 4a$ ($a =$ Cu–Cu distance) as depicted in figure 3. At $p = 0.07$, a new two-dimensional Wigner lattice order coexisting with the p(4 × 4) lattice was found and a $T_c = 26$ K was observed. This new two-dimensional Wigner lattice order has c(2 × 2) symmetry ($p_c = 1/8$, $L = 2\sqrt{2}a$) (see figure 3) and is responsible for a $T_c = 30$ K superconductivity. These new two-dimensional Wigner lattice ground states were identified by the presence of the Goldstone mode resulting from the broken translational symmetry. The $\omega = 0$ Goldstone mode ($\omega_G$) of the Wigner lattice has been shifted to $\omega_{GL} \sim 23$ cm$^{-1}$ for p(4 × 4) lattice and to $\omega_{GH} \sim 46$ cm$^{-1}$ for c(2 × 2) lattice due to the commensuration pinning of the Wigner lattice to the underlying $CuO_2$ lattice.

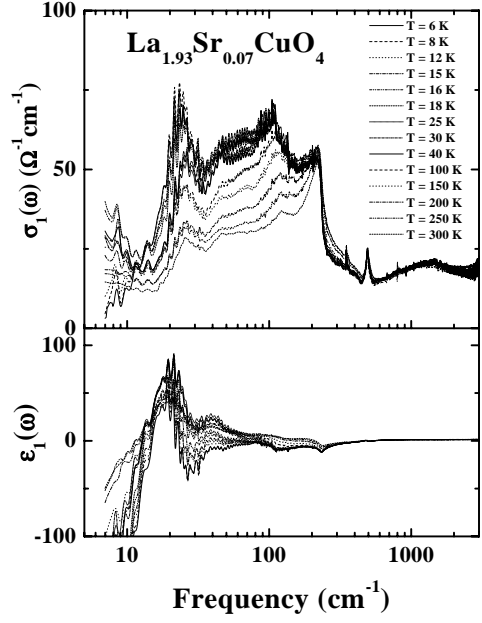

**Figure 2.** The real part of the conductivity $\sigma_1(\omega)$ (upper panel) and the corresponding real part of the dielectric function $\varepsilon_1(\omega)$ (lower panel) calculated via the Kramers–Kronig transformation of the reflectivity.

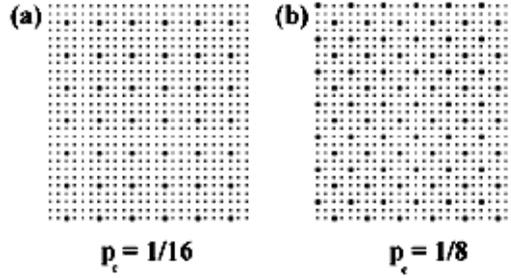

**Figure 3.** Two-dimensional square Wigner lattice diagram. (a) p($4 \times 4$) lattice and (b) c($2 \times 2$) lattice. The smaller dots represent the copper sites. The larger dots indicate the sites occupied by the electrons (or holes).

The presence of the two-dimensional Wigner lattice in cuprates plays a profound role in two important ways.

(1) Because of the on-site Coulomb repulsion at the lattice sites, the excess holes added to the lattice would occupy the upper Hubbard band if the charge motion were restricted only to the lattice sites. However, in a square lattice, an excess hole can find an energetically favoured location at the centre of the square lattice. Moreover, the excess holes ($\Delta p = p - p_c$) placed in the local potential minimum can now have access to a periodic harmonic potential in the directions to the neighbouring identical energy minima. This causes the zero-point energy of the excess holes to broaden into a band. The width is dependent on the carrier density in the band. This band (Coulomb band) is gapped by the Coulomb energy ($\Delta E_c$) above the ground state and the excess carriers are delocalized and transport charges coherently. We found that an extremely small fraction of holes (∼0.4% of $p = 0.063$, for example) participated in a coherent charge transport and superconductivity.

(2) The screening provided by the two-dimensional Wigner lattice radically influences the charge dynamics of the free carriers in the Coulomb band by providing a frequency region where the interaction energy between two electrons becomes negative. Therefore, once the long-range order of the lattice has been established, the free carriers in the Coulomb band are innately paired at temperatures below a crossover $T_o$.

In light of our previous studies, we identify the ∼24 cm$^{-1}$ mode with the Goldstone mode ($\omega_{GL}$) of the p(4 × 4) lattice formed in La$_{1.93}$Sr$_{0.07}$CuO$_4$ and the ∼100 cm$^{-1}$ peak with the transitions involving the accompanying Coulomb band ($\omega_{CL}$). This observation suggests that the Coulomb bandgap $\Delta E_c$ is of the order of ∼300 cm$^{-1}$ with the full band width, $W_c \sim 200$ cm$^{-1}$ (0.025 eV). This observed width $W_c$ gives a density of states at the Fermi energy of the order of $N(0) \sim \Delta p/(W_c/2) \sim 0.15$ states eV$^{-1}$. This gives rise to the Pauli susceptibility ($\chi_p$). Therefore, we anticipate the contribution of $\chi_p$ to the total susceptibility whenever carriers are present in the Coulomb band. We believe that the peculiar presence of $\chi_p$ even at $p = 0.05$ doping [6] is due to this factor. The presence of the Coulomb band can also be responsible for the incomplete closure of the single-particle excitation gap known as the in-plane pseudogap [7].

From the $\varepsilon_1(\omega)$ plot, it is clear that the reflectivity minimum at ∼16 cm$^{-1}$ arises from the zero crossing of $\varepsilon_1(\omega)$ and there is no sign of the related structure seen in the $\sigma_1(\omega)$. Therefore, we assign the reflectivity minimum as the plasma edge of the free carriers which gives the screened plasma frequency $\omega_p \sim 16$ cm$^{-1}$. By assuming the carrier mass is the free electron mass and using the static dielectric constant $\varepsilon_1(0) \sim 40$ estimated from $\varepsilon_1(\omega)$, we find the free carrier concentration $n_F \sim 2.3 \times 10^{17}$ carriers cm$^{-3}$ which is only 0.43% of the total doped carriers. These free carriers reside in the Coulomb band and are also responsible for the metallic behaviour indicated by the sharp upturn in $\sigma_1(\omega)$ below $\omega \sim 10$ cm$^{-1}$ and the negative $\varepsilon_1(\omega)$ for $\omega < \omega_p$. This suggests that the charge transport is coherent and nearly dissipationless as the extremely small relaxation rate (<4 cm$^{-1}$) indicates. Notice that the free carrier plasma frequency remains unchanged with $T$ (see figure 1) since the screening of the charge is also enhanced as the long range order of the Wigner lattice order develops.

A direct comparison of the $\sigma_1(\omega)$ plot with that of the previously reported La$_{1.985}$Sr$_{0.015}$CuO$_{4.032}$ sample ($p = 0.07$) is displayed in figure 4. The one-to-one correspondence of each mode in two different samples is marked. Note that even at the same hole concentration $p = 0.07$, the ∼46 and ∼70 cm$^{-1}$ modes of the La$_{1.93}$Sr$_{0.07}$CuO$_4$ sample are much weaker than the corresponding modes of the La$_{1.985}$Sr$_{0.015}$CuO$_{4.032}$ sample. At 300 K, the magnitude of the overall far-IR $\sigma_1(\omega)$ is comparable for both La$_{1.93}$Sr$_{0.07}$CuO$_4$ and La$_{1.985}$Sr$_{0.015}$CuO$_{4.032}$ samples. As $T$ is lowered, it becomes clear that the width of the $\omega_{GL}$ of the La$_{1.93}$Sr$_{0.07}$CuO$_4$ sample is about twice as broad as that of the La$_{1.985}$Sr$_{0.015}$CuO$_{4.032}$ sample while the strength of the $\omega_{CL}$ remains comparable. This implies that the correlation length of the p(4 × 4) lattice in La$_{1.93}$Sr$_{0.07}$CuO$_4$ is about half of that in La$_{1.985}$Sr$_{0.015}$CuO$_{4.032}$.

The $T$-dependence of the oscillator strength of each mode is shown in figure 5; the oscillator strength of the $\omega_{GL} \sim 24$ cm$^{-1}$ goes through two changes. There is a relatively abrupt increase in the oscillator strength at $T \sim 180$ K and another at $T \sim 16$ K. However, the $\omega_{GH} \sim 46$ cm$^{-1}$ and the Coulomb band transition $\omega_{CH}$ exhibit a peculiar $T$-dependence. First, they do not show the abrupt increase at $T \sim 200$ K as seen in the La$_{1.985}$Sr$_{0.015}$CuO$_{4.032}$ sample. Second, as the $\omega_{CH}$ mode loses its oscillator strength below $T \sim 30$ K, the oscillator strength of the $\omega_{GH}$ mode abruptly drops at T $\sim 30$ K, indicating no development of the $T_c = 30$ K SC phase collective mode. The $T \sim 180$ K jump of the $\omega_{GL}$ is an indication of the development of long-range order of the p(4 × 4) lattice at $T_o \sim 180$ K as also evidenced by the change of the dynamic mass of the lattice from $m^* \sim 75\, m_e$ at 300 K to $m^* \sim 40\, m_e$ at[3] 130 K. At the same time, the abrupt increase in the oscillator strength of the Coulomb band transition $\omega_{CL}$ at around $T \sim 200$ K suggests that more of the free carriers in the single-particle excited state of the Wigner lattice occupy the Coulomb band as temperature is lowered. However, it appears that the c(2 × 2) lattice does not develop the long-range order.

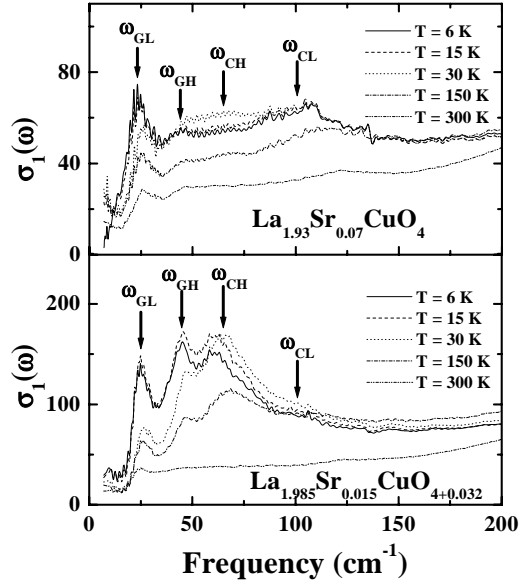

**Figure 4.** A direct comparison of the $\sigma_1(\omega)$ of $La_{1.93}Sr_{0.07}CuO_4$ with that of $La_{1.985}Sr_{0.015}CuO_{4.032}$ at selected temperatures (see the text for the notation). Notice the splitting in the peak of the Coulomb band transition in the c(2 × 2) lattice ($\omega_{CH}$) of $La_{1.985}Sr_{0.015}CuO_{4.032}$ (lower panel). One at $\sim 70$ cm$^{-1}$ is due to the transition from the Coulomb band to the single-particle excited state and the other at $\sim 62$ cm$^{-1}$ due to the transition from the ground state to the Coulomb band. The apparent splitting is caused by the relative oscillator strength changes at low temperature.

This competition is clearly seen in the $d\rho/dT$ plot in figure 6. While the $d\rho/dT$ of the $La_{1.985}Sr_{0.015}CuO_{4.032}$ sample shows a minimum at $\sim 200$ K and a maximum at $\sim 150$ K, the $d\rho/dT$ of the $La_{1.93}Sr_{0.07}CuO_4$ curve does not show such changes. The subtle changes in $\rho$ amplified by $d\rho/dT$ in $La_{1.985}Sr_{0.015}CuO_{4.032}$ indicate that as the long-range order of the both lattices develops, increases in the number of carriers in the Coulomb band and in their mobility occur and, hence, the corresponding resistivity changes result. This tendency is missing in the $d\rho/dT$ curve of $La_{1.93}Sr_{0.07}CuO_4$ owing to the short-ranged domains of the c(2 × 2) lattice embedded in the p(4 × 4) lattice. The subtle changes observed in $d\rho/dT$ of the $La_{1.985}Sr_{0.015}CuO_{4.032}$ sample have been washed out. This is because the development of the long-range order is inhibited by the frustration of holes due to the Coulomb binding of the holes with the randomly distributed immobile Sr counter-ions. In contrast, the development of long-range order of both lattices in the $La_{1.985}Sr_{0.015}CuO_{4+0.032}$ sample is enhanced by the mobile intercalated oxygen atoms. This is consistent with the fact that the correlation length of the p(4 × 4) lattice in $La_{1.93}Sr_{0.07}CuO_4$ is about half of that in $La_{1.985}Sr_{0.015}CuO_{4.032}$. We found that the short correlation length of the $c(2 \times 2)$ lattice remains unchanged and only the $p(4 \times 4)$ lattice order develops the long-range order enough to support the $T_c = 16$ K superconductivity in $La_{1.93}Sr_{0.07}CuO_4$.

---

[3] The actual $m^*$ should be larger because of the underestimation of the reflectivity for $\omega > 400$ cm$^{-1}$, which is inherent to the polycrystalline nature of the sample.

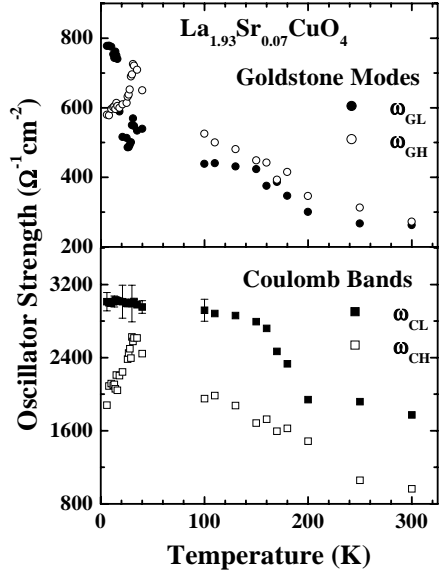

**Figure 5.** Temperature dependence of the oscillator strengths of the Goldstone modes and the oscillator strengths of the corresponding Coulomb bands. Each oscillator strength was calculated by fitting the $\sigma_1(\omega)$ with a simple symmetric Gaussian as an approximation even though non-trivial asymmetry was evident.

Two-dimensional Wigner lattice formation in a two-dimensional electron gas (2DEG) is a general phenomenon. This has been known for 2DEGs on liquid helium surfaces [8] and in Si MOSFETs [9] and $GaAs/Al_xGa_{1-x}As$ structures [10] at $B = 0$. The melting criterion for the lattice is given by considering the ratio of the Coulomb energy to the Fermi energy of the 2DEG, $r_s$ given by $r_s = 2m^*e^2/\varepsilon_o\hbar^2(\pi n_W d_\perp)^{1/2}$ where $n_W$ is the carrier density of the Wigner lattice and $d_\perp$ is the interlayer spacing. A Monte Carlo simulation for an ideally clean 2DEG found that the critical value for the Wigner crystallization of a 2DEG is $r_s = 37 \pm 5$ [11]. This limit tends to decrease as the degree of disorder increases [12]. The lattice vibration and total energy calculation [13] showed that a two-dimensional triangular lattice is favoured over a square lattice for the same reason as a simple cubic lattice is unstable against shear when the electrons interact via central forces only.

However, we propose that the two-dimensional square Wigner lattice is the ground state charge configuration in the $CuO_2$ planes away from half-filling. This is because of the symmetry of the underlying $CuO_2$ lattice and the strong electron–phonon coupling [14] which locks in the electrons to the lattice sites and further lowers the total energy. For a $p(4 \times 4)$ lattice in cuprates, with $n_W d_\perp \sim 4.4 \times 10^{13}$ carriers $cm^{-2}$ and $m^* \sim 40\,m_e$, we obtain $r_s \sim 200$, far greater than the critical $r_s$. Therefore, the lattice ground state is robust until the reduction of the dynamic mass, resulting from the increasing correlation length of the lattice, reaches the threshold beyond which a melting transition should occur. For instance, reducing the dynamic mass of the $c(2 \times 2)$ lattice ($p = 1/8$) to $m^* \sim 10\,m_e$ would give $r_s \sim 35$. Thus, at lower doping concentrations below $p = 1/8$, we anticipate the formation of a series of square lattices in the $CuO_2$ planes as $p$ starts to increase above zero. In fact, such electronic instabilities at preferred $p$ values have been found in recent Sr/O co-doping studies [15].

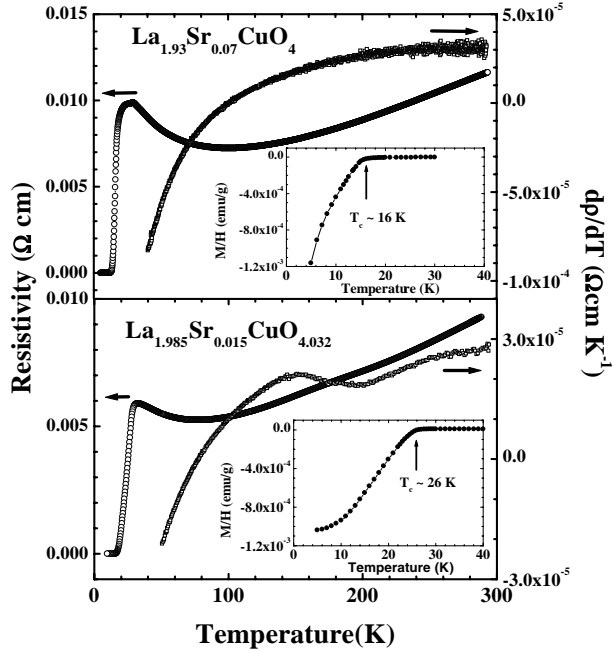

**Figure 6.** Resistivity and first derivative of the resistivity versus $T$ curves of $La_{1.93}Sr_{0.07}CuO_4$ (upper panel) and $La_{1.985}Sr_{0.015}CuO_{4.032}$ (lower panel). The inset in each panel depicts the Meissner effect. The arrow defines the onset of the SC transition temperature ($T_c$).

The gap in the Goldstone mode at a finite $\omega$ in the Wigner lattice plays an important role in the superconductivity in cuprates. This gap gives the dielectric function of the Wigner lattice $\varepsilon_1^W(q \to 0, \omega) = 1 + 4\pi n_W e^2/m^*(\omega_G^2 - \omega^2)$. Since the Coulomb band is tied to the Wigner lattice, the free carriers in the Coulomb band oscillate in phase with the underlying Wigner lattice at $\omega_G$. Then, there develops a region in $\varepsilon_1^W$ where the interaction between two electrons, $V = 4\pi e^2/q^2 \varepsilon_1^W(q, \omega)$, becomes attractive ($V < 0$) in the $q \to 0$ limit for $\omega_G < \omega < (\omega_o^2 + \omega_G^2)^{1/2}$, $\omega_o^2 \equiv 4\pi n_W e^2/m^*$. Therefore, all the free carriers in the Coulomb band are expected to form non-retarded, spin singlet pairs for $T < 180$ K which corresponds to $\omega_o \sim 130$ cm$^{-1}$ estimated from $\varepsilon_1(\omega)$. Only when $\omega > \omega_o$ (or $T > 180$ K) do these pairs of electrons dissociate through the Coulomb repulsion. Hence, we identify the crossover $T_o \sim 180$ K as the pairing temperature.

SC order requires not only the spin singlet pairing of electrons but also the long-range phase coherence of the SC order parameter, $\psi \sim \Delta e^{i\phi}$. Unlike the conventional BCS superconductors, the phase stiffness energy for cuprates is known to be much smaller [16]. Hence, the phase coherence of the pairs would develop at a $T_c$ substantially lower than the pairing $T_o$. Indeed, this point is consistent with the experimental results reported by Corson *et al* [17] and Xu *et al* [18]. Once the phase coherence has been established, there exists the phase collective mode ($\omega_\phi$) resulting from the broken gauge symmetry. By following the Nambu formalism [19] and incorporating the screened Coulomb interaction in the Wigner lattice ground state, we find $\omega_\phi \approx \omega_G/(1 + \gamma)^{1/2}$ with $\gamma \equiv n_W m_e/n_F m^*$ ($n_F$ = free carrier density) [2]. Hence, the observed abrupt increase in the oscillator strength of the $\omega_{GL} \sim 24$ cm$^{-1}$ mode is due to the development of the $\omega_\phi$ nearly at the same frequency[4] as $\omega_G$.

[4] The actual $\gamma$ should be smaller than estimated, $\gamma \sim 0.2$ for $p = 0.07$ with $m^* \sim 40\, m_e$, because of the aforementioned underestimation of $m^*$.

There is an interesting consequence of the presence of the phase collective mode in the $c$-axis charge dynamics at $T \leqslant T_c$. We propose that the presence of this phase collective mode manifests itself in the $c$-axis far-IR $\sigma_1(\omega)$ by inducing a mode arising from small fluctuations of the relative phase between the SC condensate and the Wigner lattice order. Such an idea was originally proposed by Leggett for a two-band model superconductor [20]. This mode arises from the dipole moment polarized along the $c$-axis. Therefore, a peak in the $c$-axis $\sigma_1(\omega)$ is expected ($\omega$) at $T < T_c$ as well as the longitudinal Josephson plasmon mode which solely arises from the zero crossing in the $c$-axis $\varepsilon_1$. We suggest that the unidentified SC state mode at $\sim 50$ cm$^{-1}$ in the $c$-axis $\sigma_1(\omega)$ [5] is a direct result of the coupling of the far-IR photon field to the fluctuations of the relative phase of the two condensates.

Since the formation of a series of square lattices of holes is favoured in the CuO$_2$ planes as $p$ starts to increase above zero, electrical conductivity with high mobility is expected in cuprates at elevated $T$ whenever there exist carriers in the Coulomb band of each Wigner lattice order. Ando *et al* [4] recently reported that the mobility of the conducting carriers even in 1% Sr-doped La$_2$CuO$_4$ is only a factor of three smaller than that of the optimally doped one at $T \geqslant 100$ K. Because the mobility is governed by the band width which increases with the carrier concentration in the Coulomb band, the metallic conductivity in 1% Sr-doped La$_2$CuO$_4$ can be naturally understood in our model as a direct consequence of the p(10 × 10) Wigner lattice formation and the presence of the Coulomb band at $p \sim 0.01$. However, as $T$ decreases, the excess carriers are expected to localize at a lattice site, which would lead to increasing correlation length of the lattice or to the formation of a different square lattice. Hence, in the $T \to 0$ limit, the system with $p \leqslant p_c = 1/16$ is an insulator. For a pure Sr-doped La$_2$CuO$_4$, the two-dimensional I–M transition in the Wigner lattice ground state occurs when the free carriers start to occupy the Coulomb band of the p(4 × 4) lattice for $p > p_c = 1/16$ beyond which the free carriers are paired and the system exhibits the superconductivity.

Studies of the I–M transition in cuprates with Zn doping or by applying strong magnetic field provide an important test ground for the idea of incipient pairing of electrons below $T_o$. At the I–M transition, the resistivity is expected to approach the universal critical two-dimensional resistance $\rho_c^{2D} = h/4e^2 \cong 6.5$ k$\Omega$ per CuO$_2$ layer. Indeed, this has been observed in Zn-doped cuprates where $\rho_{IM}^{2D} \sim 6.8$ k$\Omega$ per CuO$_2$ layer was found (Walker *et al* and Fuzukumi *et al* [3]). In this new type of metal–insulator transition, the Mott I–M transition criterion $k_F \ell \sim 1$ ($k_F =$ Fermi wavevector and $\ell =$ electron mean free path) is no longer valid. The observed unusual value of $k_F \ell \approx 13$ at the I–M transition boundary under strong magnetic field is a direct evidence for this type of novel I–M transition (Boebinger *et al* [3]).

We propose that the c(2 × 2) lattice starts to develop as the hole doping increases beyond $p_c = 1/16$, and at $p_c = 1/8$ the entire CuO$_2$ plane contains only the c(2 × 2) Wigner lattice. Because the Coulomb interaction energy of a square lattice per electron is $E_c \cong -0.390 e^2/L$ [13], the interaction energy difference $\delta E_c$ between p(4 × 4) and c(2 × 2) lattices is $\delta E_c = -0.404 e^2/a$ per electron. Therefore c(2 × 2) lattice formation is favoured when the Coulomb energy of the electrons in the Coulomb band of the p(4 × 4) lattice exceeds $\delta E_c$. In La$_{1.93}$Sr$_{0.07}$CuO$_4$, although both p(4 × 4) and c(2 × 2) lattices were found, the long-range order of the c(2 × 2) lattice could not develop due to the frustration of holes as observed in this work, hence no 30 K superconductivity was obtained. In this context, the observed $T_c$ ranging between 15 and 30 K for $1/16 < p < 1/8$ would be due to the proximity effect involving the c(2 × 2) lattice of $T_c = 30$ K and the p(4 × 4) lattice of $T_c = 15$ K.

Beyond $p = 1/8$, all the additional carriers are expected to occupy the Coulomb band. Hence, in our model, the free carriers with the density of $\Delta p \sim 0.035$ (or $n_F \sim 3.5 \times 10^{20}$ carriers cm$^{-3}$) undergo the SC transition at the optimal doping ($p \sim 0.16$). This gives the superfluid density fraction of $\Delta p/p = 0.035/0.16 \sim 0.21$. Therefore, our model provides a natural answer to the outstanding question of why only ∼1/5 of the total carriers participate in the superconductivity [21, 22].

It appears that the optimum doping concentration $p \sim 0.16$ is the critical point beyond which the excess carriers gain energy by occupying the $d_{3z^2-r^2}$ orbitals. In the electrochemical doping study of $La_2CuO_4$ [23], a chemical potential barrier higher than ∼1.0 eV was observed at $p \sim 0.16$. Beyond this point oxygen evolution occurs [23], hence no further doping was achieved. Based on the optical studies of $La_2CuO_4$, it has been suggested by Perkins *et al* [24] that the ∼0.5 eV absorption peak in the conductivity is due to the d–d* exciton formed by promoting a hole from $d_{x^2-y^2}$ to $d_{3z^2-r^2}$. This symmetry forbidden transition is made weakly allowed in cuprates. This peak is also seen in the photoinduced absorption studies [14] and is ∼0.5 eV lower than the calculated energy, ∼1.0 eV [24]. We suggest that the ∼0.5 eV energy difference is the d–d* exciton binding energy.

This point is further supported by reinterpreting the work of Uchida *et al* [25]. In their work, it was found on the one hand that the in-plane $\omega_p$ of the over-doped samples did not increase higher than that of $p = 0.16$ (actually it decreased slightly at $p = 0.34$) and the 0.5 eV peak in the in-plane $\sigma_1(\omega)$ loses its strength as the doping increases beyond $p = 0.16$. This observation suggests that the in-plane free carrier contribution remains fixed and the reduction in the d–d* exciton band strength directly results from the decrease of the available $d_{3z^2-r^2}$ states as the carriers occupy them. On the other hand, the Drude-like peak starts to appear in the $c$-axis conductivity for $p > 0.16$, and yet the corresponding $c$-axis plasma frequency remains small compared to the in-plane one. We believe that the systematic decrease in the resistivity (both in-plane and $c$-axis) with doping is due to the three-dimensional crossover.

## 4. Summary and conclusion

We have demonstrated that the high-$T_c$ cuprate physics has its foundation on the Wigner lattice ground state order away from half-filling. We observed a clear indication of the competition between two Wigner lattice orders; one corresponds to a $T_c = 15$ K and the other to $T_c = 30$ K. We were able to explain the metallic transport in 1% Sr-doped $La_2CuO_4$. Based on our findings, a new type of I–M transition mechanism has been established. The observed phase collective mode in the SC state provides a channel for the coupling of the $c$-axis polarized far-IR photon field to the fluctuations of the relative phase between the SC condensate and the Wigner lattice order.

## Acknowledgments

We would like to thank Zugang Li, Zheng Wu and Young Seok Song for their various technical assistance as well as sample preparations and characterizations. We also thank John Markus for his technical help to expedite our experiment. One of us (PHH) is supported by the State of Texas through the Texas Center for Superconductivity at the University of Houston.